# Do 2D material-based battery electrodes have inherently poor rate-performance?

Ruiyuan Tian, Madeleine Breshears, Dominik V Horvath and Jonathan N Coleman[*]

*School of Physics, CRANN and AMBER Research Centers, Trinity College Dublin, Dublin 2, Ireland*

*colemaj@tcd.ie (Jonathan N. Coleman); Tel: +353 (0) 1 8963859.

ABSTRACT: Two dimensional materials show great potential for use in battery electrodes and are believed to be particularly promising for high-rate applications. However, there does not seem to be much hard evidence for the superior rate-performance of 2D materials compared to non-2D materials. To examine this point, we have analyzed published rate-performance data for a wide range of 2D materials as well as non-2D materials for comparison. For each capacity-rate curve we extract parameters which quantify performance which can then be analyzed using a simple mechanistic model. Contrary to expectations, by comparing a previously-proposed figure of merit, we find 2D-based electrodes to be on average ~40 times poorer in terms of rate performance than non-2D materials. This is not due to differences in solid-state diffusion times which were similarly distributed for 2D and non-2D materials. In fact, we found the main difference between 2D and non-2D materials to be that ion mobility within the electrolyte-filled pores of the electrodes to be significantly lower for 2D materials, a situation which we attribute to their high aspect ratios.

Over the last few years 2-dimensional (2D) materials have shown huge potential for use in the number of application areas.[1,2] Some of the most promising applications have been in the field of electrochemical energy storage, particularly in the area of batteries.[3] Over the last decade, many papers have described using 2D materials, often in the form of synthesized or exfoliated nanosheets, in both lithium- and sodium-ion batteries.[4] While 2D materials have predominantly been used as active lithium or sodium storing materials, they have also been used in a number of other roles, for example as a conductive additive,[5] as a binder material,[6] and even as a separator material.[7] However, probably the most important role of 2D materials in batteries has been as active materials. While the potential for graphene to effectively store lithium was recognized very early,[8] researchers eventually began to explore transition metal dichalcogenide (TMD) nanosheets[9] before more recently branching out to explore the wider family of 2D materials to store both lithium and sodium.[4] Some to these materials have shown extremely high capacities. For example, black phosphorus-based electrodes have demonstrated sodium ion storing capacities of approximately 2500 mAh/g,[10] making it one of the most promising of all battery materials.

However, in addition to high capacity, it can be very important for electrode materials to display good rate-performance in order to facilitate fast charging or high-power delivery. Indeed, for many researchers, this is where 2D materials excel. Almost all authors claim that 2D materials tend to enable high rate-performance (in 53 out of 59 papers surveyed by us, authors claimed their 2D material displayed good rate performance, see SI table S1). Although other arguments exist (for example based on electrode morphology or conductivity, see SI table S1), the most common argument is that electrodes based on 2D materials have relatively short solid-state diffusion times, $\tau_{SSD}$, leading to fast charge/discharge (this argument has also been applied to nano-materials in general[11]). The solid-state diffusion time describes the timescale required for a Li or Na ion to diffuse within the particles of active material (AM), and is related to the diffusion length ($L_{AM}$) and diffusion coefficient ($D_{AM}$) via $\tau_{SSD} = L_{AM}^2 / D_{AM}$. Many authors argue that, for 2D materials, $\tau_{SSD}$ should be short because nanosheets tend to have small $L_{AM}$,[12-21] due to their tiny size, as well as relatively large $D_{AM}$,[22-28] because of the expectation that ion mobility within the inter-layer space would be higher than within 3D particles.

However, the evidence for this argument is relatively sparse. A large fraction of papers surveyed by us (28 out of 59 papers, see SI table S1) rely solely on reporting relatively high specific capacity (mAh/g) at relatively high specific current (mA/g) as evidence of good rate performance while the rest use this metric in combination with other experiments (31 out of 59, see SI table S1). The problem with such analysis is associated with the electrode thickness. If the electrode thickness is low, then a given specific capacity can be equivalent to a low absolute amount of charge stored, while a given specific current can be achieved for a relatively low absolute current. Achieving even theoretical capacity while inserting/extracting a small amount of charge at low current is not good evidence of impressive rate performance. In fact, real batteries need relatively thick electrodes in order to maximize charge stored as well as energy density, leading to capacity-rate tradeoffs which we have previously discussed in detail.[29] As a result, while good rate-performance might be represented by high areal capacities at high areal currents, such experiments are rarely performed.[30]

In addition, even those papers that do measure $D_{AM}$, very rarely combine it with $L_{AM}$ to estimate $\tau_{SSD}$. Even if $\tau_{SSD}$ is calculated, this number is of little use without context, i.e. what are typical values of $\tau_{SSD}$, and how big a contribution does $\tau_{SSD}$ make to the overall timescale associated with charge/discharge.

Thus, we believe that a detailed analysis of literature to assess whether or not 2D materials do indeed display good rate performance compared to non-2D materials is required. In this work we perform an extensive quantitative analysis of published rate-performance data for lithium and sodium storing electrodes based on two-dimensional materials. Using a published[31] semi-empirical equation, we fit capacity-rate data, extracting parameters which can be used to assess rate performance. Calculating a previously-proposed[31] figure of merit we find that 2D based electrodes have considerably poorer rate performance compared to non-2D materials. In addition, we find that 2D electrodes are predominantly rate-limited by diffusion effects while non-2D electrodes are limited by both diffusion and electrical effects.[31] Using a mechanistic model[31] we find that solid-state diffusion times are similar in both 2-D and non-2D electrodes. In fact, the main difference is associated with liquid diffusion within the electrolyte in the porous interior of the electrode. The high aspect ratio of 2D materials significantly reduces ion mobility, dramatically increasing liquid diffusion times. This factor is enough to significantly reduce rate performance, especially for thick electrodes.

RESULTS AND DISCUSSION

While it is well-known that the capacity of battery electrodes decreases as the rate at which they are charged/discharged is increased (see figure 1 for examples), using such data to quantify rate performance is not straightforward. Recently,[31] we proposed a semi-empirical equation which can fit capacity-rate data yielding three fit parameters which can be used to assess rate performance:

$$\frac{Q}{M} = Q_M \left[ 1 - (R\tau)^n \left( 1 - e^{-(R\tau)^{-n}} \right) \right] \qquad (1)$$

Here $Q/M$ is the measured specific capacity (mAh/g) while $R$ is the rate defined via the specific current ($I/M$) as $R = (I/M)/(Q/M)$. We note that, unlike C-rate, which is defined via the theoretical capacity, $R$ is calculated from the measured specific capacity (at a given current). In this way, $R$ is a measure of the actual charge/discharge time. Graphs of $Q/M$ vs. $R$ can be plotted (see figure 1) from typical rate data as reported in nearly all battery papers. Fitting $Q/M$ v $R$ data (see figure 1) yields $Q_M$, $\tau$ and $n$, parameters which can be used to quantify rate performance. The first parameter, $Q_M$, is the specific capacity at very low rate and represents the maximum performance of the electrode material. Perhaps more importantly, $\tau$ is a time constant associated with charge/discharge and is a measure of the rate at which $Q/M$ starts to fall off.[29, 31] This parameter is particularly important as low time constants mean fast charge/discharge and indicate good rate performance. Finally, $n$ is an exponent describing how rapidly $Q/M$ decays at high rate. Low values of $n$ indicate slow decay and so good rate performance. Diffusion limited electrodes are thought to give $n \sim 0.5$ while electrodes whose rate performance is limited by electrical properties (i.e. capacitive-limited) yield $n \sim 1$.[31] Knowledge of $\tau$ and $n$ allows a proper, quantitative assessment of the rate performance of a given electrode and comparison with other electrodes.

The aim of this paper is to assess the rate performance of battery electrodes based on 2D materials and compare their performance to other, non-2D materials. To do this we collected ~48 rate performance data sets from the literature for lithium- or sodium-storing electrodes where the active material had a predominately 2D structure.[10, 14, 16-19, 21-24, 26-28, 32-63] These data sets encompass 28 different 2D materials grouped in the following families: graphene; transition metal dichalcogenides (TMDs); other metal chalcogenides, oxides or hydroxides, MXenes and other miscellaneous materials (see figure 2). We note that not all of these materials are layered compounds with some of them (e.g. 2D LiFePO$_4$)[63] being 2D platelet-shaped nanoparticles of materials with a 3D bonding scheme. In all cases, we extracted capacity-rate data (see methods), calculated $R$ and plotted $Q/M$ v $R$. The curves were then fitted to equation 1 and $Q_M$, $\tau$ and $n$ extracted. Some examples of fits are shown in figure 1. All fits and associated data are given in the SI, figures S1-S4.

The resultant fit parameters are presented in figure 2 plotted together in pairs. Each parameter occupies a well-defined range: $0.3<n<0.8$, $10^{-3}$ h$<\tau<10$ h and 200 mAh/g$<Q_M<2000$ mAh/g. The plots in figure 2 A and B show no clear correlation between $n$ and $\tau$ or between $n$ and $Q_M$. However, figure 2C suggests a possible correlation between $\tau$ and $Q_M$ – we will discuss this in more detail below.

It has previously been shown that, in addition to impacting specific capacity, electrode thickness has a significant effect on rate performance.[31, 64] Knowledge of $\tau$, $n$ and $Q_M$ yields an excellent opportunity to assess how the rate-performance of 2D materials depends on electrode thickness. Shown in figure 3A is the time constant, $\tau$, plotted as a function of electrode thickness, $L_E$. This graph shows a roughly quadratic ($\tau \propto L_E^2$) scaling over the cohort of 2D materials as was previously observed for a broader set of electrode materials.[31] The details of this scaling will be dealt with in more detail below. Conversely, figure 3B shows no clear dependence of $n$ on electrode thickness, although over the entire thickness range the data appears to cluster around $n\sim0.5$. Again, this will be discussed in more detail below. The data for $Q_M$ versus electrode thickness is shown in figure 3C. While this data is not specifically associated with rate performance, $Q_M$ represents a good estimate of the maximum achievable capacity (at very low rate), and cannot be accurately obtained without performing rate analysis such as that outlined here. As such, it is worth a brief discussion. This graph suggests that thinner electrodes tend to display higher specific capacity, a fact that has been previously observed.[40] To test this, we normalised $Q_M$ to the theoretical specific capacity (where available) and plotted this ratio versus electrode thickness in figure 3D-E. Interestingly, we found a significant number of results with capacity well above the theoretical value (figure 3D). Although this has been previously observed, particularly for MoS$_2$,[41] this data shows that a number of other metal chalcogenides/oxides also display anomalously high capacities. Interestingly, this cohort shows no clear dependence of normalised capacity on thickness. Shown in figure 3E is normalised capacity plotted versus thickness for those data which show normal behaviour (i.e. capacity at or below the theoretical limit). Interestingly, this data set shows a clear decay of relative capacity with electrode thickness, highlighting the difficulty of maintaining high specific capacity at high electrode thickness (see ref[65] for more discussion on this topic).

In order to quantitatively assess rate performance, a Figure of Merit (FoM) is required. To achieve this, $\tau$ alone is not appropriate because it depends strongly on the electrode thickness ($L_E$),[31, 66] as shown in figure 3A. We note that both figure 3A and a previously reported literature-data analysis[31] both show an approximate scaling of $\tau \propto L_E^2$. This empirical observation is supported by a simple model recently proposed by us,[31] which relates the charge/discharge time constant to the mechanistic factors effecting rate: the RC charging time of the electrode, the timescales associated with ion diffusion and the delay time due to the electrochemical reaction:

$$\tau = L_E^2 \left[ \frac{C_{V,eff}}{2\sigma_E} + \frac{C_{V,eff}}{2\sigma_{P,E}} + \frac{1}{D_{P,E}} \right] + L_E \left[ \frac{L_S C_{V,eff}}{\sigma_{P,S}} \right] + \left[ \frac{L_S^2}{D_{P,S}} + \frac{L_{AM}^2}{D_{AM}} + t_c \right] \quad (2a)$$

Here $C_{V,eff}$ is the effective volumetric *capacitance* of the electrode, $\sigma_E$ is the out-of-plane electrical conductivity of the electrode material, $\sigma_{P,E}$ and $\sigma_{P,S}$ are the ionic conductivities of the electrolyte within the pores of the electrode and separator respectively, $D_{P,E}$ and $D_{P,S}$ are the ionic diffusion coefficients in the electrolyte within the pores of the electrode and separator respectively, while $L_S$ is the separator thickness. In addition, $L_{AM}$ is the solid-state diffusion length associated with the active particles (related to particle size); $D_{AM}$ is the solid-state Li ion diffusion coefficient within the particles such that $\tau_{SSD} = L_{AM}^2 / D_{AM}$ is the solid-state diffusion time. N.B. $D_{AM}$ is an effective value, averaged over the relevant potential and state-of-charge ranges. Finally, $t_c$ is a measure of the timescale associated with the electrochemical reaction once electron and ion combine at the active particle. The origin of each term in this equation has been explained in detail previously.[31] This equation has been shown to accurately describe a wide range of experimental data and makes predictions which are consistent with observations.[29, 31]

Equation 2a has seven terms, each representing a distinct rate-limiting factor. We have previously argued that not all of these terms are important under all circumstances.[31] For example, terms 5 and 7 represent (5) the time required for ion diffusion in the separator and (7) the timescale associated with the electrochemical reaction respectively, with both relatively unimportant under normal circumstances[31] (see SI figure S5 for further justification for neglecting these two terms). Term 1 represents the contribution to the RC charging time associated with the electrical resistance of the electrode. This term can be neglected where the electrode is conductive enough (i.e. out of plane conductivity >>1 S/m),[31] as should be the case for well-designed systems. (We accept that this will not always be the case and recommend routine out-of-plane conductivity measurements on electrodes where quantitative rate analysis is to be performed.) In addition, we note that in porous systems, both diffusivity and ionic conductivity tend to be reduced by a factor $f$, compared to that in the bulk liquid (i.e. $D_{BL}$, $\sigma_{BL}$): $f = D_{P,E}/D_{BL} = \sigma_{P,E}/\sigma_{BL}$. Finally, we note our recently reported empirical observation that $C_{V,eff}$ is directly proportional to the volumetric capacity of the electrode, $Q_V$: $C_{V,eff}/Q_V = 28$ F/mAh (N.B. $Q_V$ is the volumetric capacity at low rate: $Q_V=\rho Q_M$, where $\rho$ is the electrode density). Combining all of these observations, we can significantly simplify equation 2a, yielding:

$$\frac{\tau}{L_E^2} \approx \left[\frac{14Q_V}{\sigma_{BL}f} + \frac{1}{D_{BL}f} + \frac{28Q_V L_S/L_E}{\sigma_{P,S}} + \frac{L_{AM}^2/L_E^2}{D_{AM}}\right] \quad (2b)$$

where $Q_V$ should be expressed in mAh/m³. These remaining terms represent the contribution of the resistance of the electrolyte within the porous electrode to the RC charging time (first term), the diffusion time for ions in the electrolyte within the porous interior of the electrode (second term), the contribution to the RC time constant due to the conductivity of the electrolyte in the separator (third term), and the time associated with ion diffusion within the lithium-storing particles (fourth term). This equation is known because almost all the parameters are accessible. For a given experiment, $\tau$, $L_E$, $Q_V$ and $L_S$ should be known in all cases while we can estimate $\sigma_{BL} \approx \sigma_{P,S} \approx 1$ S/m and $D_{BL} = 3 \times 10^{-10}$ m²/s. In addition, $L_{AM}$ is related to the particle size which can be measured while, as we will see below, $f$ can be estimated. This leaves $D_{AM}$ as the only unknown in equation 2b.

We note that when the electrode thickness is large compared to the solid-state diffusion length ($L_{AM}$) and the separator thickness ($L_S$), as would be the case in real electrodes, the third and fourth terms in equation 2a will become small. This means that, especially for thick electrodes, we expect $\tau \propto L_E^2$ to be a reasonable approximation, supporting the meta data mentioned above. As a result, we have proposed that $L_E^2/\tau$ can be considered a semi-intrinsic figure of merit for rate performance in battery electrodes.[31] Large values of this FoM indicate good rate performance, consistent with relatively short charging times, even for thick electrodes.

We have combined the values of $\tau$ described above with values of electrode thickness extracted from the relevant publications (see methods) to calculate $L_E^2/\tau$ for each of the 2D-based electrodes described above. We have plotted the FoM as a histogram in Figure 4A. For this cohort of 2D materials, we find $L_E^2/\tau$ to vary between $10^{-14}$ and $10^{-11}$ m$^2$/s. The logarithmic mean was found to be $\langle \log(L_E^2 \tau^{-1}/m^2 s^{-1}) \rangle$=-12.4.

To put these numbers into context, we reproduce data for $L_E^2/\tau$ found by analysing a much wider data set of 122 results representing lithium storing materials of all types (referred to below and in the figures as "All materials"). This broader data set includes a relatively small number of 2D materials. The resultant data is plotted as a histogram in Figure 4B and shows the majority of $L_E^2/\tau$ data within the wider family of lithium storing materials to vary between $10^{-12}$ and $10^{-9}$ m$^2$/s with a logarithmic mean of $\langle \log(L_E^2 \tau^{-1}/m^2 s^{-1}) \rangle$=-10.7.

It is clear from figures 4A-B that 2D-based electrodes have a much lower FoM compared to the wider set of materials. This is a considerable difference with the shift between the distributions in Figure 4A and B indicating 2D materials to have a FoM for rate performance typically ×40 times smaller than non-2D materials. This is clear evidence that 2D-based battery electrodes have rate performance which is much poorer than battery materials in general.

To investigate why this might be, we plot the exponent, *n*, data for 2D-based electrodes reported in figure 2 as a histogram (Figure 4C). This histogram shows a single peak centred around 0.5. As indicated above, values of *n* close to 0.5 are associated with diffusion limitations. As before, we can compare this data to a histogram extracted from ref[31] which plots *n*-values from a much broader range of battery materials, of which 2D materials are only a minor component (Figure 4D). This wider set of materials shows weak peaks at *n*=0.5 (representing diffusion limitations) and *n*=1 (representing electrical limitations). However, the majority of data lies in the range 0.5<*n*<1 indicating a combination of diffusion and resistance limitations. This data suggests that while the broader set of battery materials yield electrodes which have a range of rate limiting mechanisms, 2D-based electrodes tend to be predominantly diffusion limited. We believe this result is linked to the relatively low FoMs displayed by 2D-based electrodes (Figure 4A).

We believe that the low FoMs displayed by 2D-based electrodes, coupled with the fact that rate performance appears to be diffusion limited in these materials is intrinsic to 2D materials and is linked to the electrode morphology which is associated with 2D building blocks. 2D-based electrodes consist of networks of 2D sheets separated by electrolyte-filled pores (and in most cases polymer binder and a conductive additive). Ions within the electrolyte moving through the pores must travel around nanosheets and so follow a tortuous path. This means that to travel between any two points within a pore system, ions must travel much farther than would be necessary within bulk liquid. In battery research, such reduction in ion mobility within pores is usually addressed via the Bruggeman equation which ultimately yields an expression for the ion diffusivity (and similar for conductivity) within pores: $f = P_E^{3/2}$, where $P_E$ is the electrode porosity.[67] For highly porous electrodes, $P_E$ will be ~1 while 50% porosity still yields $f$=0.35, meaning tortuosity doesn't have a dramatic impact on ion mobility.

However, this relationship strictly applies only to pseudo-spherical particles. Tortuosity should have a much bigger effect in pore systems associated with networks of 2D particles. This is the reason why nanosheets have been so successful as barrier materials.[68] A number of models have estimated the effect of tortuosity on diffusion of gasses through nanosheet networks.[69] The simplest model treats networks of aligned nanosheets and, for our purposes can be expressed as:[69]

$$f = \left[1 + (1-P_E)\frac{L}{2t}\right]^{-1} \qquad (3)$$

where $L$ and $t$ are the average nanosheet length and thickness. We note that while this expression is usually written in terms of nanosheet volume fraction ($V_f$), here it is more appropriate rewrite volume fraction in terms of electrode porosity ($V_f \approx 1-P_E$). For simplicity, we neglect the presence of binder and conductive additives.

Equation 3 is significant because, depending on the value of $L/t$, it can yield much smaller values of $f$ than found using the Bruggeman equation. For example, for nanosheet aspect ratios of L/t=200, $f$ =0.02 implying that in-pore diffusivities and ionic conductivities might be up to two orders of magnitude below bulk values in high aspect ratio nanosheet systems.

To test this, we note that equations 2a-b suggest that $\tau/L_E^2$ should scale with electrode volumetric capacity, $Q_V$, behaviour that is hinted at in figure 2C. In Figure 5A-B, we plot $\tau/L_E^2$ versus $Q_V$, for both the larger cohort representing all materials (A) as well as the narrower cohort of 2D-based electrode materials (B). In both cases, although the data sets display much scatter, a clear scaling of $\tau/L_E^2$ with $Q_V$ can be seen, with the 2D materials shifted upward compared to the broader data set due to their poorer rate performance. The nature of the scatter comes from the fact that both data sets contain electrodes of many different compositions and architectures, each with different electrode thickness, porosity, particle size and type. However, we would expect the bulk-electrolyte ion diffusivity ($D_{BL}$) and conductivity ($\sigma_{BL}$) to be similar in all cases.

The scatter makes it very difficult to quantitatively analyse the data set as a whole. However, we can get around this by considering the lower limit of each data set. In equation 2b, the first two terms are always present, simply because $\sigma_{BL}$ and $D_{BL}$ have well-defined values set by the electrolyte. However, when electrodes are thick (i.e. $L_E \gg L_S, L_{AM}$), the last two terms (representing ion flow in the separator and solid-state diffusion) can be neglected. This allows us to consider a lower limit to equation 2c which should act as a lower envelope to the data:

$$\left(\frac{\tau}{L_E^2}\right)_{Min} \approx \frac{14 Q_V}{\sigma_{BL} f} + \frac{1}{D_{BL} f} \tag{4}$$

We would expect the broad data set representing all electrode materials to contain some highly porous materials. Thus, we can model the lower bound of that data set by taking $f=1$ and assuming reasonable values of $\sigma_{BL}=1$ S/m, $D_{BL}=3\times10^{-10}$ m$^2$/s. Then, plotting equation 4 onto Figure 5A yields the grey line, which provides a reasonable match to the lower limit of the broader (grey) data set.

We have hypothesised above that the cohort of 2D materials (Figure 5B) shows poorer rate performance (ie higher $\tau/L_E^2$) because tortuosity significantly reduces $D_{P,E}$, resulting in low values of $f$. Thus, we would expect the lower bound to the 2D data in Figure 5B to be reproduced by again plotting Figure 5 but using a value of $f<1$. The blue solid line in Figure 5B is a plot of equation 4 using the same values of $\sigma_{BL}$ and $D_{BL}$ as before but taking $f=0.1$ which, according to equation 3 is associated with nanosheets with aspect ratios of ~50-60. This is a reasonable value, consistent for example with transition metal dichalcogenide nanosheets prepared by liquid exfoliation.[70]

For both the 2D and broader data sets, most of the data sits well above the lower bounds described above. Considering equation 2a, there might be two main reasons for this scatter above the lower bounds. While we would not expect $\sigma_{BL}$ and $D_{BL}$ to vary significantly over the range of commonly used electrolytes, we might expect some electrodes to have lower values of $f$ if high aspect ratio nanosheets were used. Alternatively, we might expect some variation in the fourth term in equation 2b leading to significant increases in $\tau/L_E^2$ above its lower bound. This could occur if the electrode thickness ($L_E$) were small or the solid-state diffusion time ($\tau_{SSD} = L_{AM}^2 / D_{AM}$) were large.

To assess the relative importance of these factors in determining the scatter in $\tau / L_E^2$, we plot $\tau / L_E^2$ versus $L_E$ for both the broad data set representing all materials (figure 5C) and the set of 2D materials (figure 5D). In both cases, the data shows the expected scatter but also indicates the general trend of appearing to decrease at low $L_E$ but saturate at high $L_E$. The importance of these plots is that these data sets should also be described by equation 2b. While we cannot fit the data sets as a whole because of the scatter associated with variations in electrode properties, we can, as before, consider lower, and in this case also upper, bounds to the data.

The lower bounds to both data sets occur when $Q_V$ and $L_{AM}^2 / D_{AM}$ are minimised while $f$ is maximised (we assume $\sigma_{BL}$=1 S/m, $D_{BL}$=3×10$^{-10}$ m$^2$/s). In line with figures 5 A-B, we take $f$=1 and 0.1 for "all materials" and "2D materials" respectively. We estimate the minimum values of $Q_V$ from the data spread in figures 5A-B to be ~100 and ~250 mAh/cm$^3$ for "all materials" and "2D materials" respectively. Then, using trial and error we varied the solid-state diffusion time ($L_{AM}^2 / D_{AM}$) until we got a reasonable lower bound to the data (solid lines in figure 5 C-D) This was achieved for values of $L_{AM}^2 / D_{AM}$ = 5 and 10 s for "all materials" and "2D materials" respectively.

We then followed a similar procedure to obtain upper bounds for the data sets in figure 5 C-D. We estimate the maximum values of $Q_V$ from the data spread in figures 5A-B to be ~2000 and ~3000 mAh/cm$^3$ for "all materials" and "2D materials" respectively. We then used trial and error to find maximum values of $L_{AM}^2 / D_{AM}$ = 3000 s (all) and 5000 s (2D) and minimal values of $f$=0.5 (all) and 0.1 (2D) leading to reasonable upper bounds for the data sets in figure 5 C-D) (dashed lines). We not that the minimal value of $f$=0.1 for 2D materials is a very rough estimate. As shown in the SI (figure S6), there is simply not enough literature data for 2D-based electrodes with thicknesses beyond 100 μm to properly assess the minimum value of $f$, other than that it lies somewhere between ~0.02 and 0.2. For simplicity, we take an intermediate value of $f$=0.1 and interpret the similarity of maximal and minimal values of $f$ to mean that nanosheet aspect ratio does not vary over a larger range among the 2D materials under study.

While these maximal and minimal values are of course approximate, they are instructive. An important point is that for both "all" and 2D materials, the data is consistent with a relatively small range of *f*-values. This implies that, for both data sets, the spread in $\tau/L_E^2$ is not predominately due to differences in tortuosity within the data set. For example, in the 2D data set, this means that variations in aspect ratio or morphology do not have a large impact on $\tau/L_E^2$. However, for both data sets the maximum and minimum values of solid-state diffusion time ($L_{AM}^2/D_{AM}$) are significantly different, with the maximum value being ~500 times larger than the minimum value in each case. This implies that, for both data sets, most of the spread in $\tau/L_E^2$ is due to variations in solid-state diffusion time. This could be down to either variation in particle size (represented by $L_{AM}$) or solid-state diffusion coefficient ($D_{AM}$). Importantly, both lower and upper bounds for $\tau_{SSD} = L_{AM}^2/D_{AM}$ are similar between the broader set representing all materials and the 2D materials data set. This implies that, contrary to expectations, 2D materials do not have any significant advantage when it comes to solid-state diffusion as might be expected due to the potential for high diffusion coefficients within the interlayer space.

This means that the main differences between the data sets in figures 5 C and D is the fact that while the broader set of materials have values of *f* roughly in the range 0.5-1, 2D materials have much smaller values of *f* in the region of 0.1. This difference indicates that the differences in rate behaviour between the two cohorts is predominately due to tortuosity in 2D based electrodes which is associated with their high aspect ratio and leads to reduced ionic mobility. Although this is a problem that is inherent to 2D materials, it may be resolvable simply by using 2D materials with reduced aspect ratio. Ironically, the best 2D materials for rate performance in batteries may be those which are poorly exfoliated.

Finally, the similarity between the upper and lower bounds of *f* for both "all materials" and 2D materials data sets allows us to use this data to examine the timescale associated with diffusion of ions within the active particles. The solid-state diffusion time is related to both particle size and solid diffusion coefficient via $\tau_{SSD} = L_{AM}^2/D_{AM}$. Equation 2b can be rearranged to give $\tau_{SSD}$ once a number of other parameters are known:

$$\tau_{SSD} = \tau - L_E^2 \left[ \frac{14Q_V}{\sigma_{BL} f} + \frac{1}{D_{BL} f} + \frac{28Q_V L_S/L_E}{\sigma_{P,S}} \right] \tag{5}$$

Again, $\tau$, $L_E$, $L_S$ and $Q_V$ are known in all cases while we can estimate $\sigma_{BL} \approx \sigma_{P,S} \approx 1$ S/m and $D_{BL}=3\times10^{-10}$ m$^2$/s. Thus, within this approximation, $f$ is the only unknown. However as shown in figure 5 C-D, the upper and lower bounds of $f$ are similar for each data set, allowing us to approximate $f$ as constant in each case with average values of 0.75 (all materials) and 0.1 (2D materials). This allows us to estimate $\tau_{SSD}$ to a reasonable degree of accuracy for all sample within both cohorts. These data are plotted as histograms in figures 6 A-B for "all materials" and 2D materials respectively. We will justify the accuracy of this data below. In each case, we see relatively broad distributions in the range 1 s<$\tau_{SSD}$<$10^4$ s. Both these distributions have similar logarithmic means of $\langle \log(\tau_{SSD}/s) \rangle$=2.5 and 2.2 for "all materials" and 2D materials respectively. Again, this indicates that, on average, 2D materials have only slightly lower solid-state diffusion times compares to other materials.

It is also useful to calculate the ratio of solid-state diffusion time to the overall time constant associated with charge/discharge (i.e. $\tau_{SSD}/\tau$). This ratio is of interest as it indicates how significant the contribution of solid-state diffusion is to the overall time constant. We have plotted this ratio versus electrode thickness in figure 6 C-D for "all materials" (C) and 2D materials (D). Because the data set is so extensive for "all materials", figure 6C shows a very well-defined trend. For low electrode thicknesses, $\tau \sim \tau_{SSD}$, meaning rate performance is dominated by solid-state diffusion within particles as might be expected. However, as electrode thicknesses increase past ~10 µm, $\tau_{SSD}/\tau$ begins to fall. This is because, for thicker electrodes, factors such as the time associated with diffusion of ions within the electrolyte-filled porous interior of the electrode, become non negligible and eventually begin to dominate. Figure 6C implies that for electrodes thicker than a few hundred microns, solid-state diffusion is no longer dominant in most systems. Roughly the same behaviour can be seen for the 2D materials in figure 6D, although the trend is not quite as clear, probably because there are fewer data points in figure 6D compared to figure 6C. However, for 2D materials the $\tau_{SSD}/\tau$ data begins to fall off at lower values of $L_E$ compared to figure 6C. This indicates that factors such as liquid diffusion in the porous electrode becomes dominant earlier (i.e. at lower thicknesses) in 2D systems. This is completely consistent with the fact that $f$ (and so $\sigma_{BL}$ and $D_{BL}$) is considerably lower for 2D materials resulting in reduced ion mobility within the electrolyte filled pores.

We can test the accuracy of the $\tau_{SSD}$ values for 2D materials by independently estimating the solid-state diffusion time via $\tau_{SSD} = L_{AM}^2 / D_{AM}$. However, care must be taken here because, strictly speaking, $L_{AM}$ a characteristic length associated with diffusion within particles, rather than the actual particle size. Jiang et al.[ref71] have proposed that, for spherical particles, $L_{AM}$ is one third of the particle radius. Assuming this relationship can be applied to nanosheets, then $L_{AM}$ is roughly one sixth of the nanosheet lateral size (*L*). This yields $\tau_{SSD}=(L/6)^2/D_{AM}$. To calculate $\tau_{SSD}$, we obtained solid-state diffusion coefficients ($D_{AM}$) from the literature for as many 2D materials as possible (see SI table 3). In addition, where possible we extracted nanosheet lateral sizes (*L*) from the papers in question. However, in many cases, mean nanosheet sizes are not given, forcing us to estimate sizes from TEM/SEM images while in a few cases, it was impossible to estimate *L*. Thus we accept that values of $\tau_{SSD}$ contained in this way will have great scope for error, partly because of the crudeness of the size measurements and partly because sample-to-sample differences may make the $D_{AM}$ values inaccurate. Nevertheless, we plot values of $\tau_{SSD}$ estimated in this way versus values estimated using equation 5 in figure 6E. Notwithstanding the uncertainty, we find reasonably good agreement between values of $\tau_{SSD}$ calculated by both methods with most data points sitting near the dashed line representing y=x. Such agreement implies that equation 5 can successfully yield $\tau_{SSD}$ from standard rate performance data once *f* is estimated. We note that for non-2D materials, *f* can always be estimated from the porosity via the Bruggeman equation[67], while for 2D materials *f* can be estimated from the combination of porosity and aspect ratio (equation 3). The exception to this good agreement is the data for graphene which was calculated using $D_{AM}=10^{-14}$ m$^2$/s, a well-established value for graphite.[72] The data in figure 6E implies that the actual effective solid-state diffusion coefficient for the graphene used here is significantly lower than this value. However, it is worth noting that graphite comes in many forms with reported diffusion coefficients varying over four orders of magnitude so perhaps this disagreement is not surprising.[73]

If we accept that equation 5 is reasonably accurate, then the $\tau_{SSD}$ data obtained from it can be used to estimate the solid-state diffusion coefficient via $D_{AM}=(L/6)^2/\tau_{SSD}$ once the particle size, *L* is known. Using nanosheet sizes, *L*, estimated from each paper as described above, we calculated $D_{AM}$ for 35 different 2D data sets encompassing both Na and Li ion batteries. These values were then ordered from lowest to highest and allocated a sample number running from 1 for the smallest value to 30 for the largest. This data is plotted in figure 6F as sample number versus $D_{AM}$. As shown in ref[74], when plotted this way, the data approximates the cumulative distribution function for the data set (in this case, the distribution of solid-state diffusion coefficients for 2D materials). Figure 6F implies that 2D materials tend to have Li and Na ion solid-state diffusion coefficients predominately in the range $10^{-18}$-$10^{-13}$ m$^2$/s. In fact, this range is quite similar to that found[75] for non-2D materials where, for example, LiFePO$_4$ might have a diffusion coefficient as low as $10^{-18}$ m$^2$/s while graphite or NMC display values as high as $10^{-13}$ m$^2$/s. Thus, this work raises questions over the conventional wisdom that 2D materials display advantages associated with fast solid-state diffusion.

While there are many interesting things to note in figure 6F, we note only a few. First, three of the graphene samples show very low $D_{AM}$, below $3\times10^{-18}$ m$^2$/s and probably much lower than expected. In each case, these graphene samples were made by liquid phase exfoliation[47, 48] which involves sonication of graphite in solvents. We hypothesise that sonochemistry may have occurred at the nanosheet edges which may hamper the entry of Li ions into the basal plane.

Around the middle of the distribution, there is a cluster of Li storing TMDs between sample numbers 15 and 25 which the model gives diffusion coefficients in the range 2-6$\times10^{-17}$ m$^2$/s. This cluster contains three MoS$_2$ and one TiS$_2$ samples. According the literature, these materials have Li solid-state diffusion coefficients of $15\times10^{-17}$ m$^2$/s (ref[76]) and $4\times10^{-17}$ m$^2$/s (ref[77]) respectively, in reasonably good agreement with the model predictions. Finally, it is worth noting that the three highest $D_{AM}$ values of 6-9$\times10^{-14}$ m$^2$/s all come from VS$_2$ based electrodes. These values are all close to the value of ~$10^{-13}$ m$^2$/s reported for Na ion transport in VS$_2$.[49]

Conclusions

In conclusion, we have extracted 48 capacity-rate data sets representing 25 different 2D materials from the literature, taking care to also extract the electrode thickness, $L_E$ in each case. These were fitted using a semi-empirical equation yielding three fit parameters: the low rate specific capacity, $Q_M$, the charge/discharge time constant, $\tau$, and the high-rate exponent, $n$, parameters which can be used to assess the rate performance. This 2D data set was compared to a similar, previously published data set representing a wide range of predominately non-2D materials. We found that 2D materials tended to have longer time constants than other materials suggestive of poorer rate behaviour. By comparing a previously proposed figure of merit for rate performance ($L_E^2/\tau$) we found that 2D materials are on average ×40 times poorer than other materials. Analysis of $n$ shows 2D materials to be predominately rate-limited by diffusive effects while other materials tend to display both diffusive and electrical limitations. Using a simple model to analyse the dependence of $\tau/L_E^2$ on both $L_E$ and the low rate volumetric capacity, $Q_V$, we found the range of solid-state diffusion times to be similar for both 2D and non-2D materials. However, we found the ionic mobility within the electrolyte-filled porous interior of the electrode to be significantly lower for 2D materials compared to non-2D materials. We believe this to be a consequence of the morphology of 2D-based electrodes where ions are forced to follow tortious paths as they travel through the electrodes.

Methods

Capacity-rate data were extracted from published papers using the "digitizer" function in Origin. Charge/discharge rate is generally expressed via current or C-rate. These parameters were converted to rate, $R$, via the equations given in ref [31]. All fitting was performed using Origin software (here we used Origin version 2015-2018) via the "Nonlinear Curve Fit" function, according to equation 1. Care must be taken in fitting, with the best results obtained by fitting the log of capacity versus rate as described in ref [31]. All fits and associated data are given in the supplementary information. We note that the vast majority of published papers do not give enough information to properly analyse rate data. While active material loading (mg/cm$^2$) and proportions of active material versus binder and conductive additive are *usually* given, electrode thickness is rarely explicitly mentioned in battery papers. This is unfortunate as equation 2A makes clear that thickness has a critical impact on rate performance. In order to facilitate rate analysis, we were forced to estimate electrode thickness in most cases (see SI). We did this considering: A) the total mass loading and mass fraction of active material; B) the densities of active material and binder/additive combination and C) the electrode porosity. The parameters marked A are usually given in papers – where they are not, analysis is impossible. The parameters marked B can almost always be estimated with reasonable accuracy. However, the porosity (C) is very rarely given even though it is clearly critical for rate performance (as it impacts in ion diffusion in the electrolyte within the porous interior of the electrode). In most cases, we were forced to estimate the porosity. Unless otherwise stated, we set the electrode porosity at P=0.5. This is justifiable for 2D-based films as measurements have shown them to have porosity close to this value.[78] Assuming the actually porosity to lie in the range 0.4-0.6 yields a porosity error of 20%. Assuming the mass loading error is ~10% then yields an error in electrode thickness of roughly 30% which is acceptable given the very broad range over which $L_E^2/\tau$ is distributed. The data sets in figures 3-5 representing a wide range of material types (labelled "all") is taken directly from ref [31] and comprises the data labelled "cohort I, standard lithium ion electrodes".

**Acknowledgments:** All authors acknowledge the SFI-funded AMBER research centre (SFI/12/RC/2278) and Nokia for support. JNC thanks Science Foundation Ireland (SFI, 11/PI/1087) and the Graphene Flagship (grant agreement n°785219) for funding.

Figures

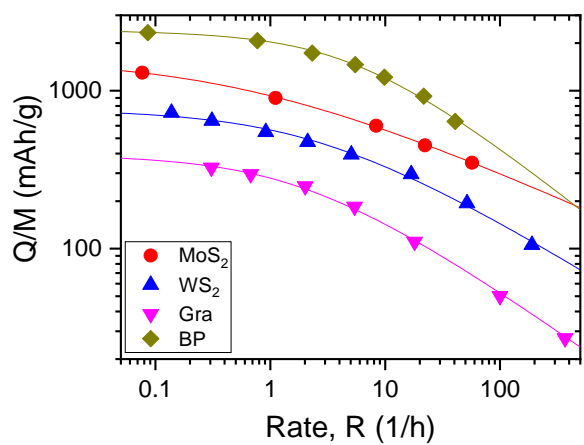

Figure 1: Examples of capacity versus rate data for battery electrodes based on 2D materials. The lines are fits to equation 1. Data are taken from the following papers: $MoS_2$ [18]; $WS_2$ [23]; Graphene [45]; black phosphorous [10]. All fits are shown in the SI.

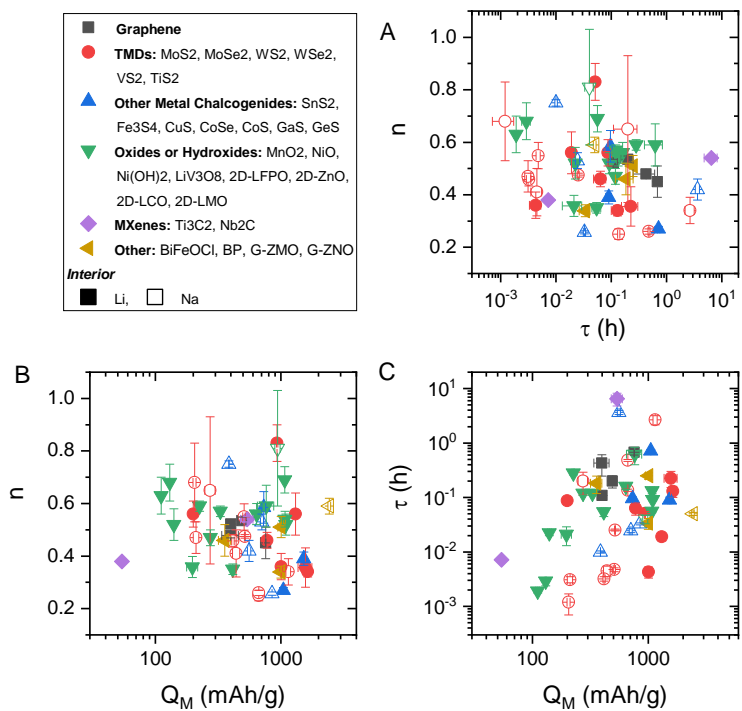

Figure 2: Fit parameters, obtained by fitting 48 data sets obtained from the literature (see SI, figures S1-4) to equation 1. These data sets encompass 25 different 2D materials grouped in the following families: graphene; transition metal dichalcogenides (TMDs); other metal chalcogenides, oxides or hydroxides, MXenes and other miscellaneous materials. The individual materials making up the families are given in the legend. Closed symbols represent lithium ion batteries while open symbols represent sodium ion batteries. In A-C, these parameters are plotted against each other in three different combinations.

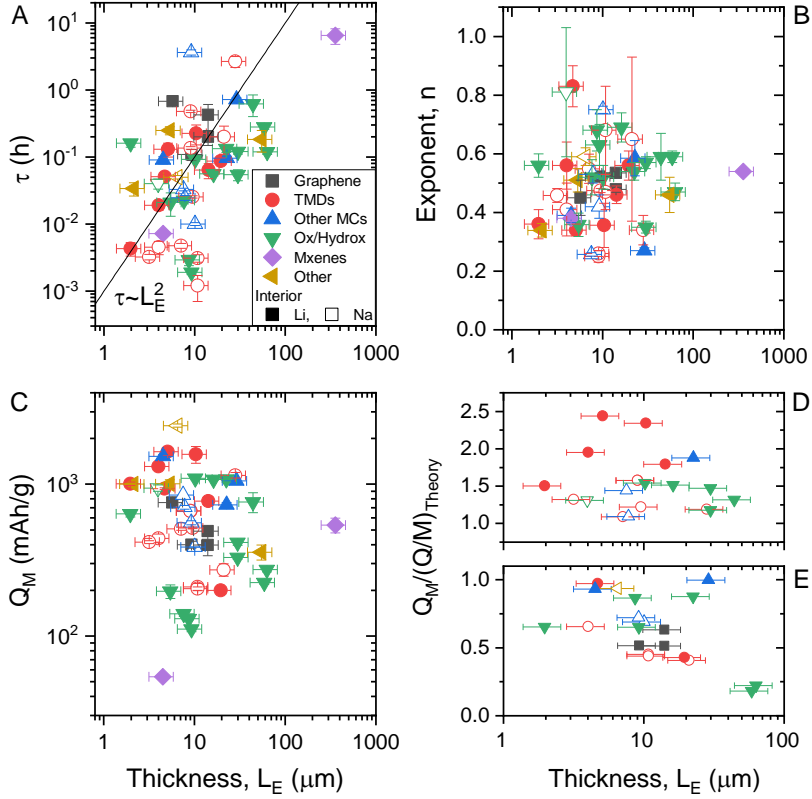

Figure 3: Thickness dependence of fit parameters. A-C) Thickness dependence of (A) time constant, $\tau$, (B) exponent, $n$, and (C) low rate capacity, $Q_M$. D-E) $Q_M$ normalised to theoretical specific capacity for (D) those materials showing anomalous behaviour ($Q_M > (Q/M)_{Theory}$) and (E) those materials showing normal behaviour ($Q_M \leq (Q/M)_{Theory}$). Closed symbols represent lithium ion batteries while open symbols represent sodium ion batteries. The legend in A applies to all panels.

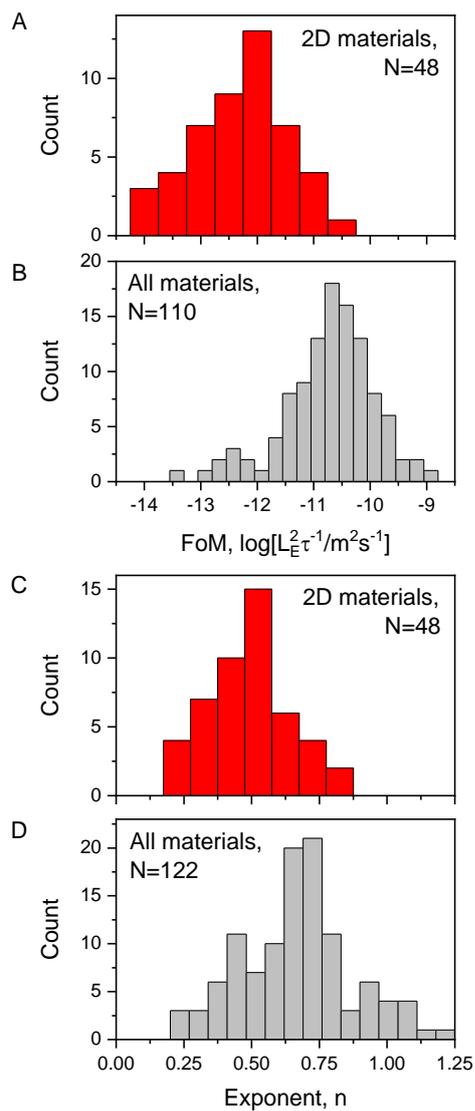

Figure 4: Histograms comparing the figure of merit for rate performance (A, B) and rate exponent (C, D) between 2D materials-based electrodes (A, C) and a wider data set including electrodes fabricated from many material types (B, D).

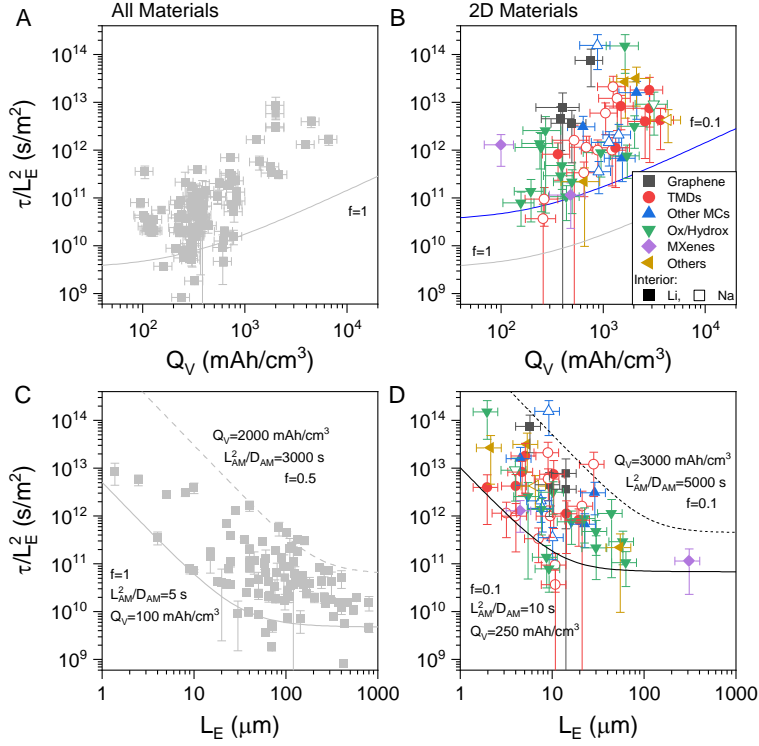

Figure 5: A-B) Time constant normalised to square of electrode thickness, $\tau/L_E^2$, plotted versus volumetric capacity of electrode, $Q_V$, for electrodes based on (A) "all materials" and (B) 2D materials. We note that $\tau/L_E^2$ is the inverse of the figure of metric plotted in Figure 4 such that low values of $\tau/L_E^2$ represent better rate performance. The lines in A-B are plots of equation 4 for $f=1$ (gray) and $f=0.1$ (blue), taking $L_S=25$ μm, $\sigma_{BL}=\sigma_{P,S}=1$ S/m, $D_{BL}=3\times10^{-10}$ m²/s. This equation simulates the lower limit of $\tau/L_E^2$, an approximation which is valid for thick electrodes or short diffusion times. Then, rate performance is limited only by ionic motion in the electrolyte within the porous interior of the electrode. The grey line in A-B represents the situation where ionic conductivity and diffusivity in the pores are equal to their values in bulk electrolyte. The blue line in (B) represent the situation when these parameters are reduced by a factor of $f=0.1$ relative to bulk liquid. C-D) Plots of $\tau/L_E^2$ versus electrode thickness, $L_E$, for electrodes based on (C) "all materials" and (D) 2D materials. The lines represent upper (dashed) and lower (solid) limits of $\tau/L_E^2$ for a given $L_E$. These were found by plotting equation 2b taking $\sigma_{BL}=1$ S/m, $D_{BL}=3\times10^{-10}$ m²/s and using the parameters in the panels. N.B. the legend in B) also applies to D). Closed symbols represent lithium ion batteries while open symbols represent sodium ion batteries.

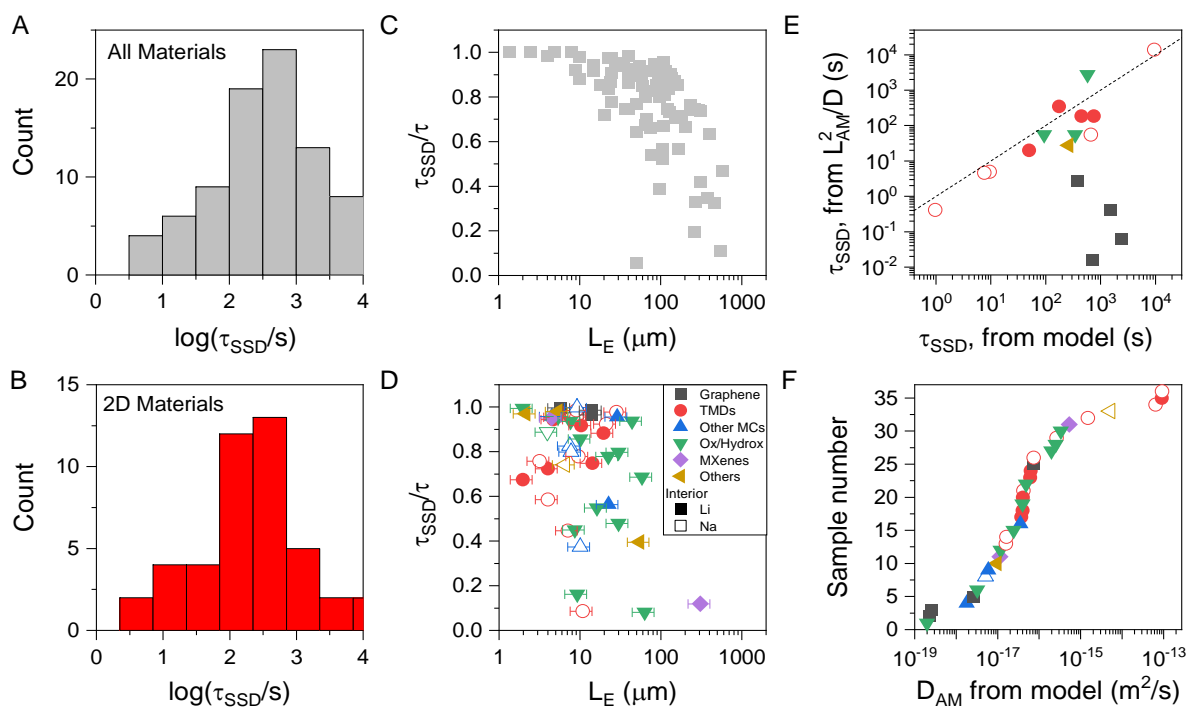

Figure 6: A-B) Histograms showing estimated solid-state diffusion time ($\tau_{SSD}=L_{AM}^2/D_{AM}$) for "all materials" (A) and 2D materials (B). C-D) Ratio of solid-state diffusion time to time constant associated with charge/discharge for "all materials" (C) and 2D materials (D). E) Solid-state diffusion time for 2D materials found using the estimated particle size and the diffusion coefficient extracted from the literature plotted versus the extracted from model (using eq 5). The dashed line represents y=x. F) Solid-state diffusion coefficient of 2D materials extracted from $\tau_{SSD}$ (model) and the diffusion coefficient (literature) plotted in ascending order. Plotted this way, this graph approximately represents the cumulative distribution of diffusion coefficients. N.B. the legend in D) also applies to E-F). Closed symbols represent lithium ion batteries while open symbols represent sodium ion batteries.